\documentclass[12pt]{iopart}
\usepackage[normalem]{ulem}
\usepackage{geometry}
\usepackage{cite}
\usepackage{tikz}
\usepackage{pgfplots}
\usetikzlibrary{decorations.pathmorphing}
\usetikzlibrary{decorations.markings}
\usetikzlibrary{decorations.text}
\definecolor{cytop}{HTML}{0002D7}
\definecolor{sio}{HTML}{0027B2}
\definecolor{alo}{HTML}{008652}
\definecolor{mgo}{HTML}{009643}
\definecolor{hbn}{HTML}{005F7A}
\definecolor{hbn2}{HTML}{00D404}
\definecolor{hbn3}{HTML}{00993F}
\definecolor{laalo}{HTML}{D50300}
\usepackage{widetext}

\begin{document}

\title[]{Engineering of Neutral Excitons and Exciton Complexes in Transition Metal Dichalcogenide Monolayers through External Dielectric Screening}

\author{Sven Borghardt$^1$, Jhih-Sian Tu$^1$, Florian Winkler$^{2,3}$, J\"urgen Schubert$^1$, Willi Zander$^1$, Kristj\'an Lesson$^4$, Beata E. Kardyna\l{}$^1$}
\address{$ˆ1$ Peter Gr\"unberg Institute 9 (PGI-9), Forschungszentrum J\"ulich, DE-52425 J\"ulich, Germany}
\address{$ˆ2$ Ernst Ruska-Centre for Microscopy and Spectroscopy with Electrons (ER-C), Forschungszentrum J\"ulich, DE-52425 J\"ulich, Germany}
\address{$ˆ3$ Peter Gr\"unberg Institute 5 (PGI-5), Forschungszentrum J\"ulich, DE-52425 J\"ulich, Germany}
\address{$ˆ4$ Innovation Center Iceland, Keldnaholti, IS-112 Reykjavik, Iceland}

\ead{b.kardynal@fz-juelich.de}

\vspace{10pt}
\begin{indented}
\item[] May 2017
\end{indented}

\begin{abstract}  
In order to fully exploit the potential of transition metal dichalcogenide monolayers (TMD-MLs), the well-controlled creation of atomically sharp lateral heterojunctions within
  these materials is highly desirable. A promising approach to create such heterojunctions is the local
  modulation of the electronic structure of an intrinsic TMD-ML via dielectric screening induced by its surrounding materials.
  For the realization of this non-invasive approach, an in-depth understanding of such dielectric effects is required.
  We report on 
  the modulations of excitonic transitions in TMD-MLs through the effect of dielectric environments including low-k and high-k dielectric materials. 
  We present absolute tuning ranges as large as $37$~meV for the optical band gaps of WSe$_2$ and MoSe$_2$ MLs and relative
  tuning ranges on the order of $30\%$ for the binding energies of neutral excitons in WSe$_2$ MLs. The findings suggest the possibility to reduce
  the electronic band gap of WSe$_2$ MLs by 
  $120$~meV, paving the way towards dielectrically defined lateral heterojunctions.
\end{abstract}

\vspace{2pc}

\ioptwocol

\section{Introduction}

Since the discovery of graphene the class of atomically thin
two-dimensional materials has been growing continuously\cite{novoselov2004,li2014,vogt2012,davila2014}.
An exciting expansion of this class was done when transition metal dichalcogenide monolayers (TMD-MLs) 
were added to this material class \cite{mak2010}. 
In contrast to the semimetal graphene, TMD-MLs with the stoichiometric formula MX$_2$, where M and X denote a transition metal atom (Mo, W, ...) and a chalcogen atom (S, Se, ...), respectively, are semiconductors with direct band gaps at the $K$-points of the hexagonal Brillouin zone with energies in the range of visible light \cite{rasmussen2015}.
Additionally, the heavy metal ions, the reduced Coulomb screening in two dimensions and the lack of inversion symmetry in these materials
result in a unique combination of giant exciton binding energies, large spin-orbit coupling, as well as a coupling between the spin and valley degrees of freedom at the two inequivalent $K$-valleys that can be accessed optically \cite{ugeda2014,cao2012,xiao2012}.

Although there have been numerous reports on electronic and optoelectronic nanodevices made from TMD-MLs
such as single-layer transistors, light-emitting diodes and photodetectors \cite{radisavljevic2011,ross2014,lopez2013}, there is a strong need to develop methods to create junctions.
In analogy to well established semiconductor technology, where heterojunctions are the building blocks of most electronic and optoelectronic devices (e.g. laser diodes, HEMT transistors),
the reliable creation of lateral heterojunctions within TMD-MLs would enable further development of next generation optoelectronic devices. 

\begin{figure}
\centering
 \includegraphics{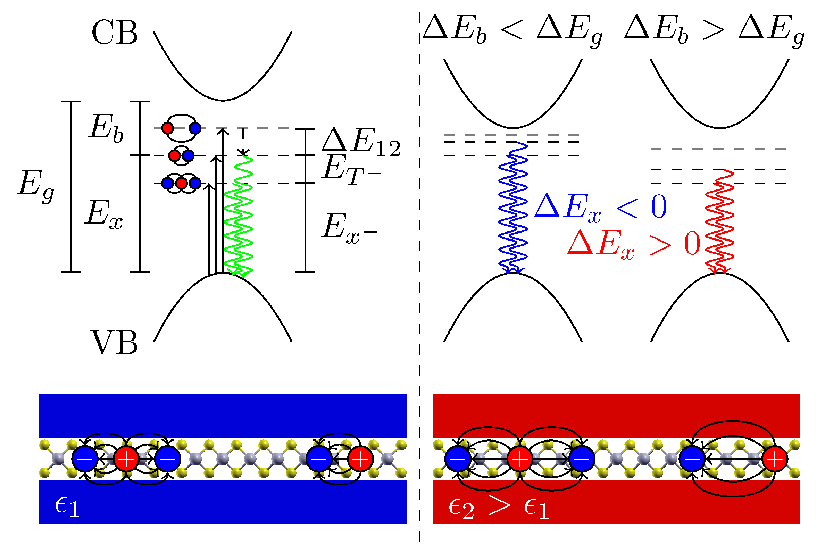}
 \caption{\label{fig:bands}Schematic diagram of externally induced dielectric screening effects
 on the electronic and optical band gaps at the $K$-point
 of TMD-MLs. Solid arrows, dashed arrows and waved arrows represent optical absorption processes,
 non-radiative relaxation processes and radiative recombination processes, respectively. 
 }
\end{figure}

In analogy to three dimensional semiconductors where heterostructures are formed by interfacing materials of different chemical composition, lateral heterostructures of TMDs have been developed by introducing laterally varying stoichiometry \cite{huang2014,mahjouri2015}.
An alternative approach, which is unique to low dimensional materials,
relies on the local modulation of screening of the ionic Coulomb potentials, as well as the Coulomb interaction between charge carriers,
within the ML\cite{roesner2016,ryou2016} through a local variation of the relative permittivity of the ML's direct environment, i.e. its substrate and cover material. 

Fig.~\ref{fig:bands} depicts how externally induced screening  changes both the electronic and optical band gaps
of a TMD-ML.
Both the electronic band gap, $E_g$, and the binding energy of the neutral exciton, $E_b$, in the monolayer are expected to decrease with an increasing relative permittivity of the surrounding dielectric \cite{roesner2016,ryou2016,latini2015,kylanpaa2015}. 
The sign of the change of the optical band gap, $E_x = E_g - E_b$, depends on the relative change of these two quantities and can be probed optically by measuring the energy of free, neutral excitons. 
In addition to neutral excitons, charged exciton states (trions) with binding energies, $E_{T^\pm} = E_x-E_{x^\pm}$, on the order of $30$~meV 
have been observed in TMD-MLs\cite{ross2013,mak2012,wang2014,plechinger2015}.
Similarly to the binding energies of  neutral excitons, the binding energies of trions in ML's are expected to decrease 
with an increasing background relative permittivity\cite{kylanpaa2015}. Negatively charged trions are expected 	to be present  in unbiased TMD-MLs, which are known
to be unintentionally n-doped\cite{ross2013,mak2012,wang2014,plechinger2015}.

\begin{figure}[t]
\centering
 \includegraphics{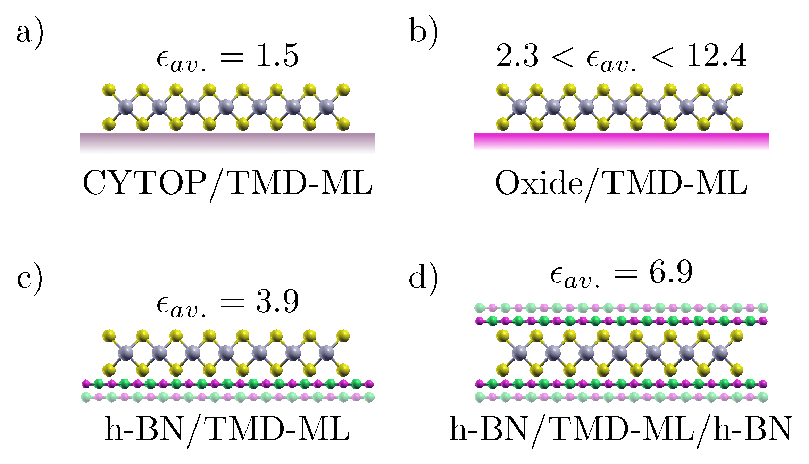}
 \caption{\label{fig:environments} Symbolic diagram of the structures studied in this work, together with the average relative permittivity
 of the substrate and cover material in each structure.}
\end{figure}

A first experimental evidence for the tunability of excitonic properties in TMD-MLs through environment-induced dielectric screening
has been the report of a room temperature blue shift of $E_x$ and $E_{x^-}$ in  MoS$_2$ MLs with an increasing relative permittivity of the non-ionic liquid with which the samples
were covered\cite{lin2014}. More recently, 
an increase of the exciton Bohr radius in WSe$_2$ MLs with an increasing background relative permittivity {of the surrounding material of the ML} has been observed\cite{stier2016}.
In addition, a decrease of the exciton binding energy and the electronic band gap has been observed for WS$_2$ MLs stacked with graphene\cite{raja2017}.

In this work, we investigate different types of dielectrics as potential components of lateral heterostructure.
We employ low temperature $\mu$-photoluminescence, as well as reflectance measurements to study MoSe$_2$ and WSe$_2$ MLs. 
Our results show a red-shift of the optical band gaps, as well as a reduction of the binding energies of both the neutral exciton and
the trion,
with an increasing average relative permittivity of the ML's surrounding. Estimates on the resulting {changes of the} electronic band gaps are made 
from the measured optical band gaps and exciton binding energies.

\section{Experiments}

Mechanically exfoliated WSe$_2$ and MoSe$_2$ MLs have been prepared on different substrates.
The substrates were required to be flat in order to minimize 
the strain and the number of defects introduced in the TMD-MLs. Only transparent
wide band gap materials were used. The subsrates considered in this work can be separated into  three types 
(fig.~\ref{fig:environments} a)-c))  according to their surface properties. 
The first type of substrate, namely CYTOP, is a low-k fluoropolymer, which is known for its chemical inertness and hydrophobicity. 
It is amorphous and transparent across a wide wavelength range and it features a very low relative permittivity for the entire frequency range.
The second type are stable, transparent oxide substrates.  Oxides offer a wide range of k-values but they are known to be hydrophilic. Hexagonal boron-nitride (h-BN)
is a wide band gap layered material that can form van-der-Waals heterostructures with TMD-MLs. Similarly to CYTOP, h-BN is hydrophobic. 
A TMD-ML encapsulated by two hexagonal h-BN flakes was also included as a fourth dielectric environment (fig.~\ref{fig:environments} d)).

On a theoretical level, the effect of dielectric screening on excitonic properties is complex and non-analytic. Especially, an open
question remains in which frequency domains the screening occurs.
In our analysis, for the sake of simplicity, we identify the surrounding materials by their static dielectric constant. This represents a simplification,
however, we emphasize that it does not change the conclusions from this work. A more detailed discussion of this point can be found in the Supporting Information.
The average nominal relative permittivity of the environment, $\epsilon_{av.} = \frac{1}{2}\left(\epsilon_{top}+\epsilon_{bottom}\right)$,
ranged from $1.5$ for a TMD-ML on CYTOP 
to $12.4$ for a TMD-ML on LaAlO$_3$.  
All the layers in contact with the MLs
were at least several nanometers thick and, as a consequence, 
only the layers in direct contact with the TMD-ML are needed to be included in the analysis.

$\mu$-photoluminescence maps, as well as optical reflectance measurements of all structures were acquired at $11$~K.
While the former method only probes the emission energies of the ground states
of neutral excitons and trions, absorption energies of the ground state, as well as of excited states, of the neutral exciton can be probed by 
the latter \cite{chernikov2014,hill2015}.
We further analyze the impact of
charging induced by the substrate or the cover material as the reason for the observed effects.

\begin{figure}
 \centering
 \includegraphics{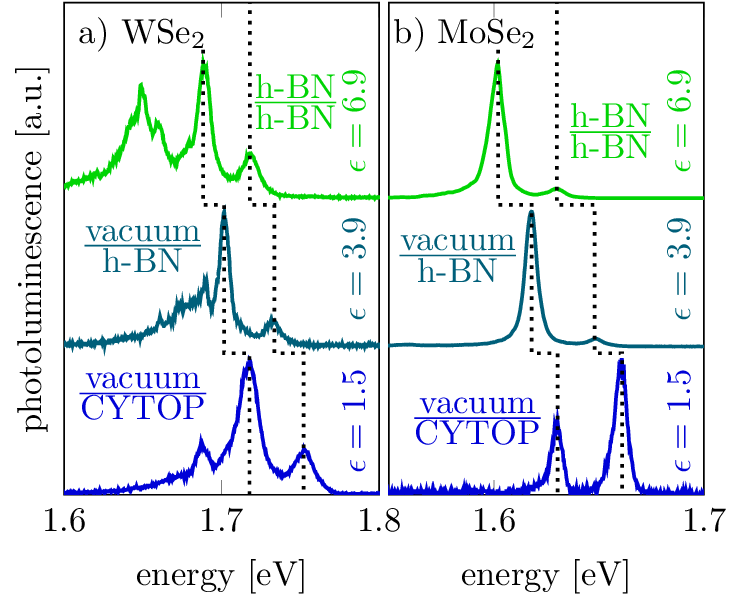}
 \caption{\label{fig:spectra_mose2}Representative photoluminescence spectra for WSe$_2$ (a) and MoSe$_2$ (b) MLs in different
 dielectric environments. The highest and second highest energy peak in each spectrum represent the neutral exciton and trion emission, respectively. For WSe$_2$,
 a band of defects can be observed at the low energy end of the spectra.
 The dotted black lines connect the neutral exciton and trion emission peaks of each spectrum, as a guidance to the eye.}
\end{figure}

\begin{figure*}
\centering
 \includegraphics{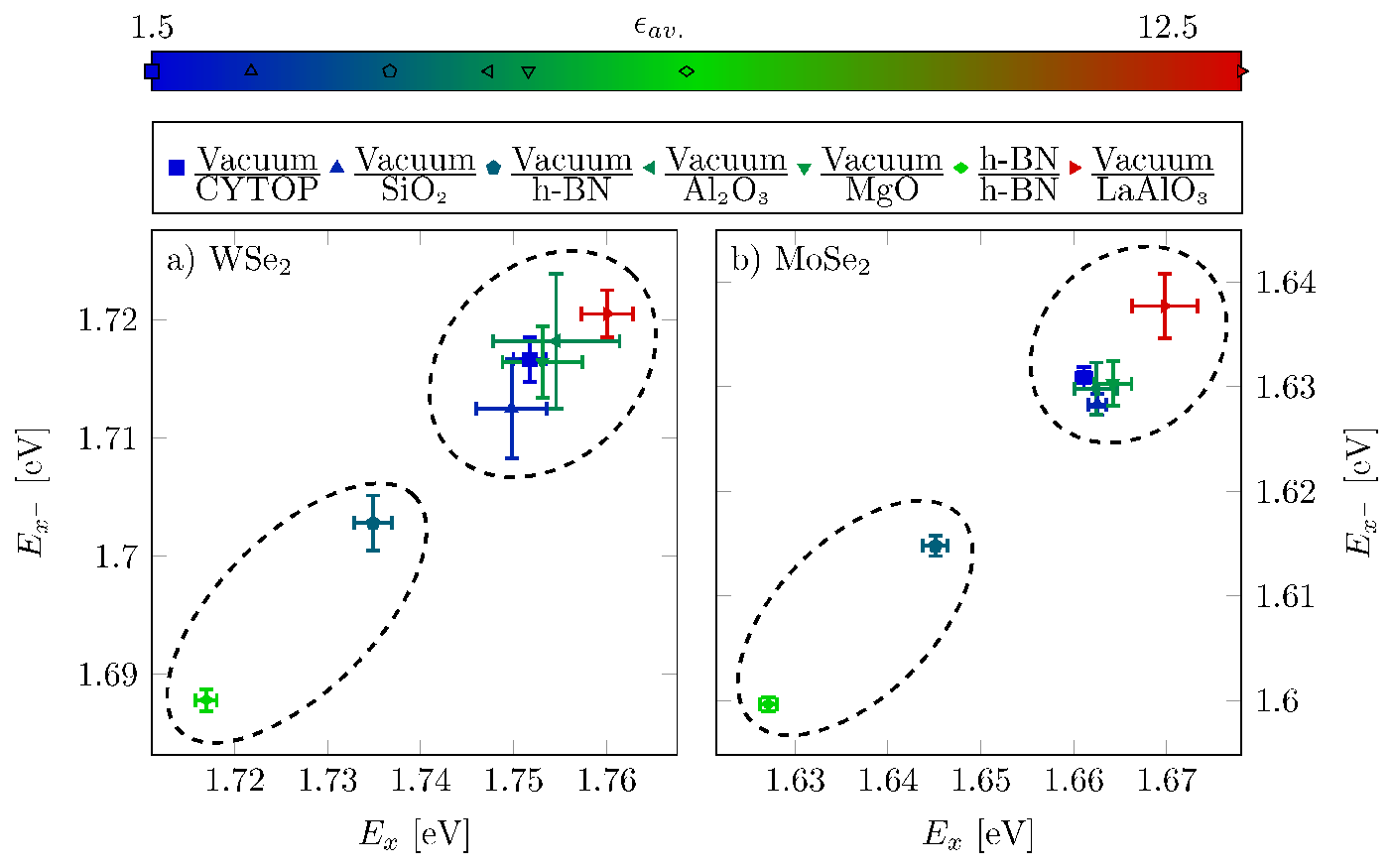}
 \caption{\label{fig:exctri}Emission energies of the neutral exciton and the trion in  WSe$_2$ (a) and  MoSe$_2$ (b) MLs for different dielectric environments.
 The legend shows the substrate and the cover material below and above the horizontal line, respectively. The color of the plotted data represents $\epsilon_{av.}$.
 The emission energies at each sample position have been obtained by hyperspectral fitting using a purpose-developed software (see Supporting Information) and the 
 mean values and error bars are obtained by a pixel-by-pixel averaging over the sample area.}
\end{figure*}

\section{Results}

The emission energies of the neutral exciton and the trion of all structures shown in figure \ref{fig:environments} have been measured
in $\mu$-photoluminescence experiments.
Figure \ref{fig:spectra_mose2} shows photoluminescence spectra for three representative structures.  The spectra contain signal from neutral exciton and trion recombination, as well as defect emission in the case of WSe$_2$ MLs. The energy of these peaks shows a systematic shift with $\epsilon_{av.}$.  
Figure \ref{fig:exctri},
summarizes the emission energies for both the neutral exciton and the trion for all measured structures.  The data points are averaged over large areas in order to account for spatial variations of the peak positions. 
The trends in behaviour of 
WSe$_2$ and MoSe$_2$ MLs in different environments are very similar. 
The results for both TMD-MLs on SiO$_2$, which is the most commonly used
substrate, agree well with reports in the literature\cite{jones2013,ross2013}.
There is a clear correlation in the positions of neutral exciton and trion peaks but there is no clear correlation with the nominal $\epsilon_{av.}$.
Irrespective of the nominal relative permittivity of the substrate, the emission energies for MLs on CYTOP and
for MLs on oxide substrates are very similar to each other and cluster around $\left(E_x,E_{x^-}\right) = \left(1.754\textrm{eV},1.717\textrm{eV}\right)$ 
for WSe$_2$ and at 
$\left(E_x,E_{x^-}\right) = \left(1.664\textrm{eV},1.631\textrm{eV}\right)$ for MoSe$_2$.
We attribute this effect to the presence of water on the
 hydrophilic oxide surface preventing
a direct contact between the substrate material and the TMD-MLs. Hexagonal ice has a relative permittivity below $2$\cite{kobayashl1982} and, thus, similar to that of CYTOP.

Even though the relative permittivity of h-BN is lower than that of some oxides,
the emission energies of MLs on a h-BN substrate, as well as of MLs encapsulated in h-BN, show clearly observable red-shifts with respect to this cluster.
For the neutral exciton, the red-shifts with respect to a ML on CYTOP 
are $17$~meV ($16$~meV) and $35$~meV ($34$~meV) for WSe$_2$ (MoSe$_2$) MLs on h-BN and  encapsulated with h-BN, respectively.
For the trion, the red-shifts  with respect to a ML on CYTOP are $14$~meV ($16$~meV) and $29$~meV ($31$~meV) for WSe$_2$ (MoSe$_2$) MLs on h-BN and encapsulated with h-BN, respectively. 
It is well known that contamination is highly mobile in mechanically stacked van-der-Waals heterostructures clustering into so-called bubbles
leaving behind large areas of clean interfaces (Supporting Info, fig. 1)\cite{haigh2012}. Since this contamination covers only a very small fraction of the sample area,
the photoluminescence emission from the clean interfaces is expected to dominate the measured signal.

The observed red-shifts for the h-BN structures imply that the electronic band gap  is undergoing
a larger absolute change than the exciton binding energies
 when increasing $\epsilon_{av.}$.
These findings are in contrast to previously reported observations in room temperature
measurements of MoS$_2$ MLs in dielectric non-ionic liquids\cite{lin2014}, but agree well with recent reflectance studies of TMD-MLs
stacked with graphene\cite{raja2017}. 

\begin{figure}[t]
 \centering
 \includegraphics{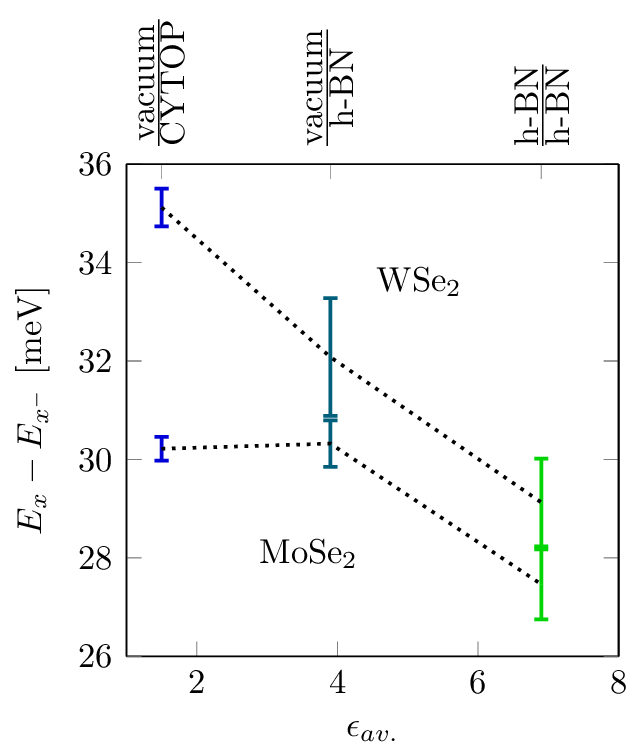}
 \caption{\label{fig:trions}Trion binding energies in WSe$_2$  and MoSe$_2$  monolayers, as a function of $\epsilon_{av.}$. 
 The binding energies at each sample position have been obtained by hyperspectral fitting using a purpose-developed software (see Supporting Information) and the 
 mean values and error bars are obtained by a pixel-by-pixel averaging over the sample area.}
\end{figure}

Figure \ref{fig:trions} shows the evolution of the trion binding energy, determined from the difference between neutral exciton and trion energies, 
as a function of $\epsilon_{av.}$, for MoSe$_2$ and WSe$_2$. 
For WSe$_2$ the trion binding energy decreases monotonically from $35$~meV  for a ML on CYTOP
to $29$~meV for a ML encapsulated in h-BN. This result is in good qualitative agreement with theoretical 
predictions \cite{kylanpaa2015}. In contrast, the trion binding energy of a MoSe$_2$ ML is nearly identical for a
ML on CYTOP and a ML on h-BN. However, similarly to WSe$_2$, the trion binding energy of MoSe$_2$ decreases from 
$30$~meV for a ML on h-BN
to $27$~meV for a ML encapsulated between two h-BN flakes. The small change between the trion binding energy
of a MoSe$_2$ ML on CYTOP and a MoSe$_2$ ML on h-BN is still under investigation. A possible explanation may be a positive polarity of the trion for a ML on CYTOP,
which, having a lower binding energy than negatively charged trions would cancel out the effect of the reduced screening \cite{jones2013}.
An indication for a different background charge for the ML on CYTOP is the different exciton-trion ratio observed in this structure (fig.~\ref{fig:spectra_mose2}).
The trion binding energies are determined as the average difference between the exciton and trion emission energies and 
the standard deviations shown in figure \ref{fig:trions} are up to five times lower than those determined for the exciton and trion emission energies themselves
(fig. \ref{fig:exctri}). This lower spatial fluctuation of the trion binding energy, indicates the robustness of excitonic binding energies
against perturbations such as strain, as expected from theoretical studies \cite{shi2013}. Trion binding energies could thus be potentially used as a sensitive probe
of surface state of transition metal dichalcogenide monolayers, providing a non-invasive alternative to the use of water droplets \cite{chow2015}.

\begin{figure*}
\centering
 \includegraphics{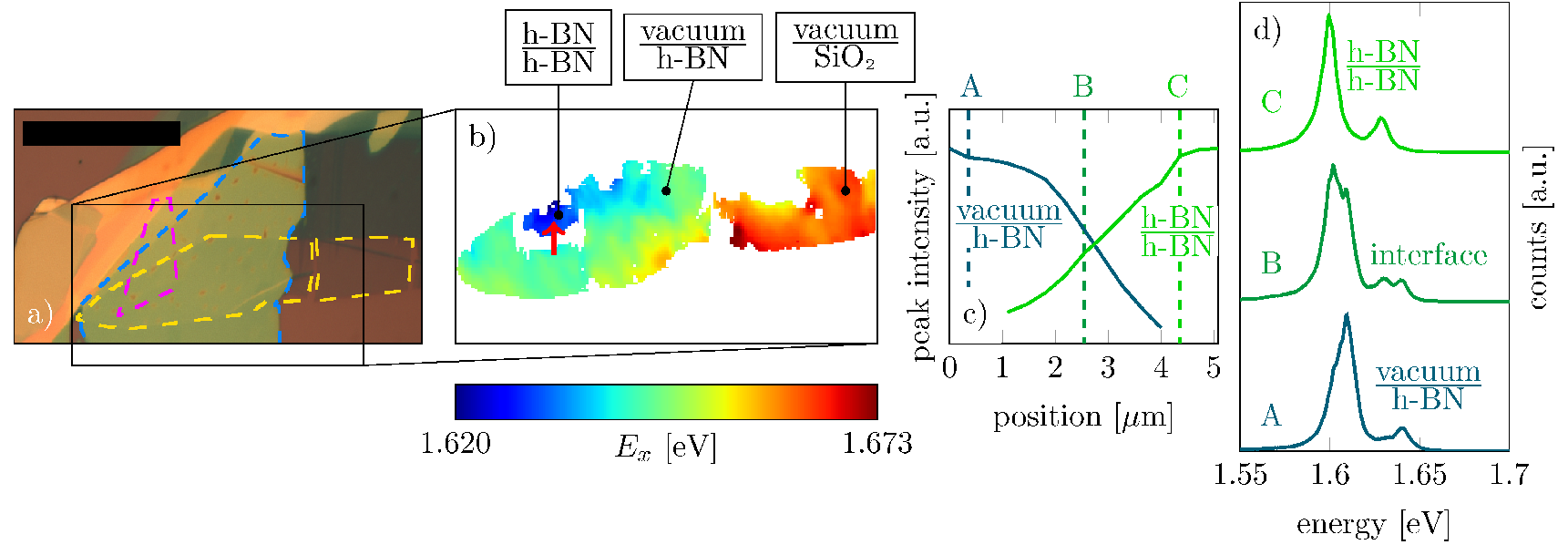}
 \caption{\label{fig:plmap} MoSe$_2$ monolayer in different dielectric environments.
 a) Optical micrograph of the TMD-ML (yellow line), which is partly lying on a SiO$_2$ substrate and partly lying
 on a h-BN flake (blue line). Furthermore, a part of the latter sample area is covered with another h-BN flake (magenta line).
 The scale bar is $20$~$\mu$m.
 b) Spatial distribution of the exciton emission energy of the TMD-ML.
 Note that some pixels needed to be removed from the analysis when they showed disturbed photoluminescence
 spectra or when they probed two dielectric environments at the same time. 
 c) Normalized intensities for the neutral exciton emission from the ML on h-BN and the ML encapsulated by h-BN along the
 a path marked by the red arrow in b). d) Representative spectra along the same path.}
\end{figure*}

Changes of the optical band gap of TMD-MLs with spatially modulated $\epsilon_{av.}$ are shown in figure \ref{fig:plmap} for 
a single MoSe$_2$ ML. The sample is prepared by placing the ML on an interface between SiO$_2$ and hBN and partially covering it with a hBN layer.
While small variations of $E_x$ within a region of constant $\epsilon_{av.}$ are present, the correlation of $E_x$ with the local $\epsilon_{av.}$
of the TMD-ML is clear. The residual variations within each region are most likely due to strain variations, which are known to change
the ML's electronic band gap\cite{johari2012}. A line profile across the boundary, that separates the part of the monolayer 
encapsulated in h-BN from the uncovered part on h-BN is shown in figure~\ref{fig:plmap}~c). In the line profile,
it can be seen that the change of the optical band gap is observed over a distance of approximately $1.5$~$\mu$m, which is 
the size of the laser spot in our measurements. 
Consequently, the band gap change is much more abrupt than the radius of the laser spot, which agrees well
with theoretical predictions of the transition occuring across only a few crystal unit cells\cite{roesner2016}. 
A representative spectrum from the interface between the two areas is shown in figure~\ref{fig:plmap}~d).
It is a clear superposition of the peaks characteristics of each of the two areas.
A line profile crossing the boundary
between the ML area on h-BN and the ML area on SiO$_2$ cannot be analyzed, as there is no smooth transition of the photoluminescence signal observable. 
The height step between the two areas, which introduces strain or defects in the ML is most likely responsible for the distortions.  The comparison between the two discussed interfaces demonstrates that the formation of well-defined, abrupt lateral heterostructures requires a change of the relative permittivity on a flat substrate.

Screening of Coulomb interactions at high carrier densities in TMD-MLs can lead to shifts in the exciton energies \cite{steinhoff2014}.  Such charges can be unintentionally introduced, for example,
via substrate doping\cite{sercombe2013}. 
In order to evaluate the possible role of doping 
 we performed gate dependent photoluminescence experiments by
 contacting the device shown in fig. \ref{fig:plmap} with graphene as a contact. 
The results of the gate dependent photoluminescence experiments are shown in the Supporting Information (Supporting Information, Figure 2).
It is apparent that, although the emission energies
of both the neutral exciton and the trion are undergoing small changes if the gate voltage is changed, they are much smaller than the effect of the dielectric environment.
In addition, different binding energies for negatively and positively charged trions can be seen, supporting the suggestion
that the measured trion binding energy for a MoSe$_2$ ML on CYTOP can be explained by a positive polarity of the trion in this structure.

\begin{figure}[t]
 \centering
 \includegraphics{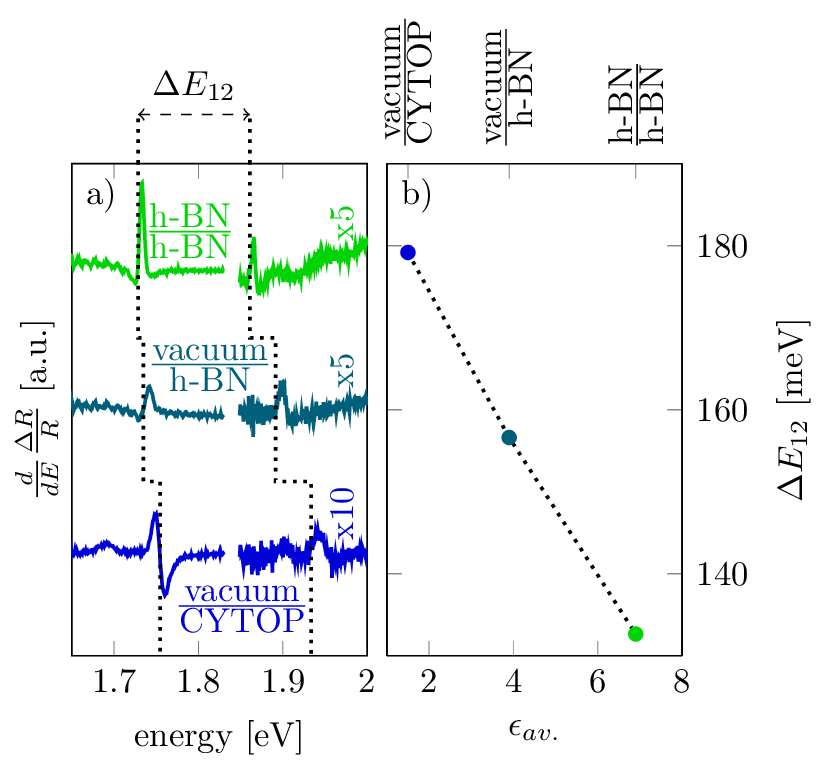}
 \caption{\label{fig:reflectance}Reflectance contrast measurements for WSe$_2$ MLs in environments with different relative permittivity. a)
 shows the energy derivative of the reflectance contrast. The dotted lines connect the ground state and the first excited state of the neutral exciton
 of each spectrum, as a guidance to the eye. The positions of the ground state and excited state were determined by optimizing the 
 energy derivative of the reflection contrast calculated using the transfer matrix method\cite{byrnes2012}. 
 b) Energy splitting of the ground state and the first excited
 state of the neutral exciton, as a function of $\epsilon_{av.}$.}
\end{figure}

So far, the discussion has been limited to photoluminescence experiments, the results of which are of importance mainly for the design of new ultra-thin 
light sources with a tailored emission energy. However, 
the electronic band gap is also affected by the dielectric screening and understanding of the interplay between the two values may be of importance for trion 
confinement, as well as confinement in electronic devices.
For this reason, we performed optical reflectance measurements which
probe Rydberg series of the neutral exciton. The energy splittings in the Rydberg series are related to the exciton binding energy, the latter combined with optical band gap yielding an estimate of the electronic band gap\cite{chernikov2014}.
The reflectance measurements were limited to WSe$_2$ MLs.
The excited states
of the neutral exciton overlap with the B-exciton in MoSe$_2$ MLs and, thus, cannot be measured\cite{wang2015}.

Due to the optical transparency of TMD-MLs, the effect of light reflection from the ML constitutes as small part of the total reflections from the sample.
For this reason, reflectance contrast measurements need to be performed on such structures. The reflectance contrast is then
given by
$ \frac{\Delta R}{R} = \frac{R_{ML}-R_{ref.}}{R_{ref.}}$,
where $R_{ML}$ and $R_{ref.}$ denote the reflectance from the ML region and the reflectance from a reference region
without a ML structure, respectively. In order to make the features of the excited exciton states more pronounced, it is convenient to
plot the energy derivative of the reflectance contrast, $\frac{d}{dE} \frac{\Delta R}{R}$\cite{chernikov2014}.

Figure \ref{fig:reflectance} shows the energy derivative of the reflectance contrast
of WSe$_2$ MLs for three measured samples  together
with the evolution of the energy splitting between the ground state and the first excited state of the neutral exciton, $\Delta E_{12}$,
as a function of $\epsilon_{av.}$. 
As in the photoluminescence experiments, the optical band gap is red-shifted. In addition,
it is apparent that $\Delta E_{12}$, and hence $E_b$, is decreasing
with an increasing relative permittivity of the surrounding material of the ML. 
$\Delta E_{12}$ decreases monotonically from approximately $180$~meV for a WSe$_2$ ML
on CYTOP to almost $130$~meV for a WSe$_2$ ML encapsulated by two h-BN flakes. Theoretical quasi two-dimensional calculations predict a relative change of $25\%$ for
the exciton binding energy when moving from a MoS$_2$ ML on h-BN to a MoS$_2$ ML encapsulated with h-BN, which agrees well with the measured
relative change in $\Delta E_{12}$ of $16\%$\cite{latini2015}.

For MoS$_2$ and WS$_2$ MLs on SiO$_2$, it has been shown that $E_b \approx 2\Delta E_{12}$, deviating from a classic hydrogenic model in three-dimensional semiconductors
with $E_b = \frac{9}{8} \Delta E_{12}$\cite{chernikov2014,hill2015}.
With the assumption that this scaling factor is independent
of the dielectric surrounding of the TMD-ML, we estimate the absolute reduction in the exciton binding energy to be on the order of $90$~meV when replacing the CYTOP substrate with h-BN encapsulation.
A decrease of $120$~meV, which is more than two times larger than the thermal energy at room temperature is obtained by combining the measured shift in the optical band gap with the reduction of the electron binding energy.

\section{Conclusions}

We have shown that h-BN is a suitable cladding material to tune the optical band gap of TMD-MLs whereas
oxide substrates, irrespective of their relative permittivisties, do not lead to a significant change in the optical band gaps of TMD-MLs.

A reduction of the optical band gaps of MoSe$_2$ and WSe$_2$ MLs by $37$~meV was measured between the uncovered ML placed on a CYTOP substrate 
and a ML encapsulated  in h-BN.  In addition, a relative change of the exciton binding energy for WSe$_2$ MLs of approximately $30\%$ was indirectly measured, which leads to an estimated
reduction of the electronic band gap of WSe$_2$ MLs by $120$~meV, when comparing uncovered 
MLs on a CYTOP substrate to a ML encapsulated in h-BN.

The energies of optical transition obtained in this work have been probed on the time scale of the exciton lifetime.  
It would be interesting to compare our results for the electronic band gap with scanning tunneling microscopy experiments, which probe the electronic band gap
in the static limit.

We have shown that formation of optical lateral heterostructure purely by local variation of screening in three dimensions will require a control of the substrate flatness as well as  substrate induced charge state.  The latter is important to avoid energy transfer between excitons and trions of different parts of the heterostructure. 

In the heterojunction studied in this
work (see fig. \ref{fig:plmap}), the change from one region to another region was abrupt on the scale of the probe beam. Therefore, techniques with a better spatial
resolution need to be developed in order to study dielectric effects on smaller length scales.

Our findings pave the way towards the creation of controlled lateral heterostructures, which are required as building blocks for next
generation electronic and optoelectronic devices. 

\section{Acknowledgements}

S.\ B. would like to thank Tony Heinz, Alexey Chernikov and Archana Raja for fruitful discussions and helpful advices.

\section{References}

\bibliography{references}

\newpage
\newgeometry{left=0cm,bottom=0cm,top=0cm,right=0cm}

\begin{figure*}
\centering
 \includegraphics{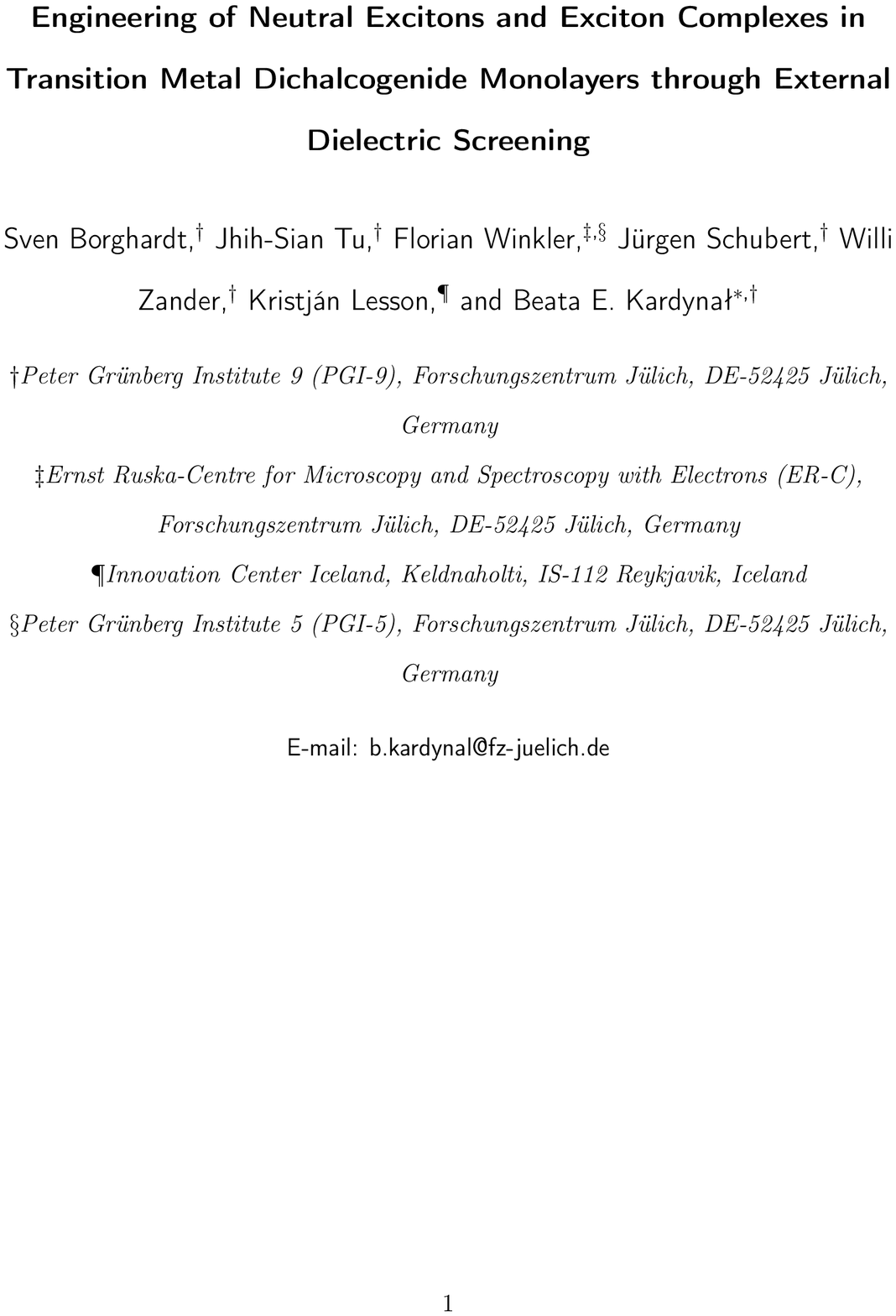}
 \caption{}
\end{figure*}

\newpage 

\begin{figure*}
\centering
 \includegraphics{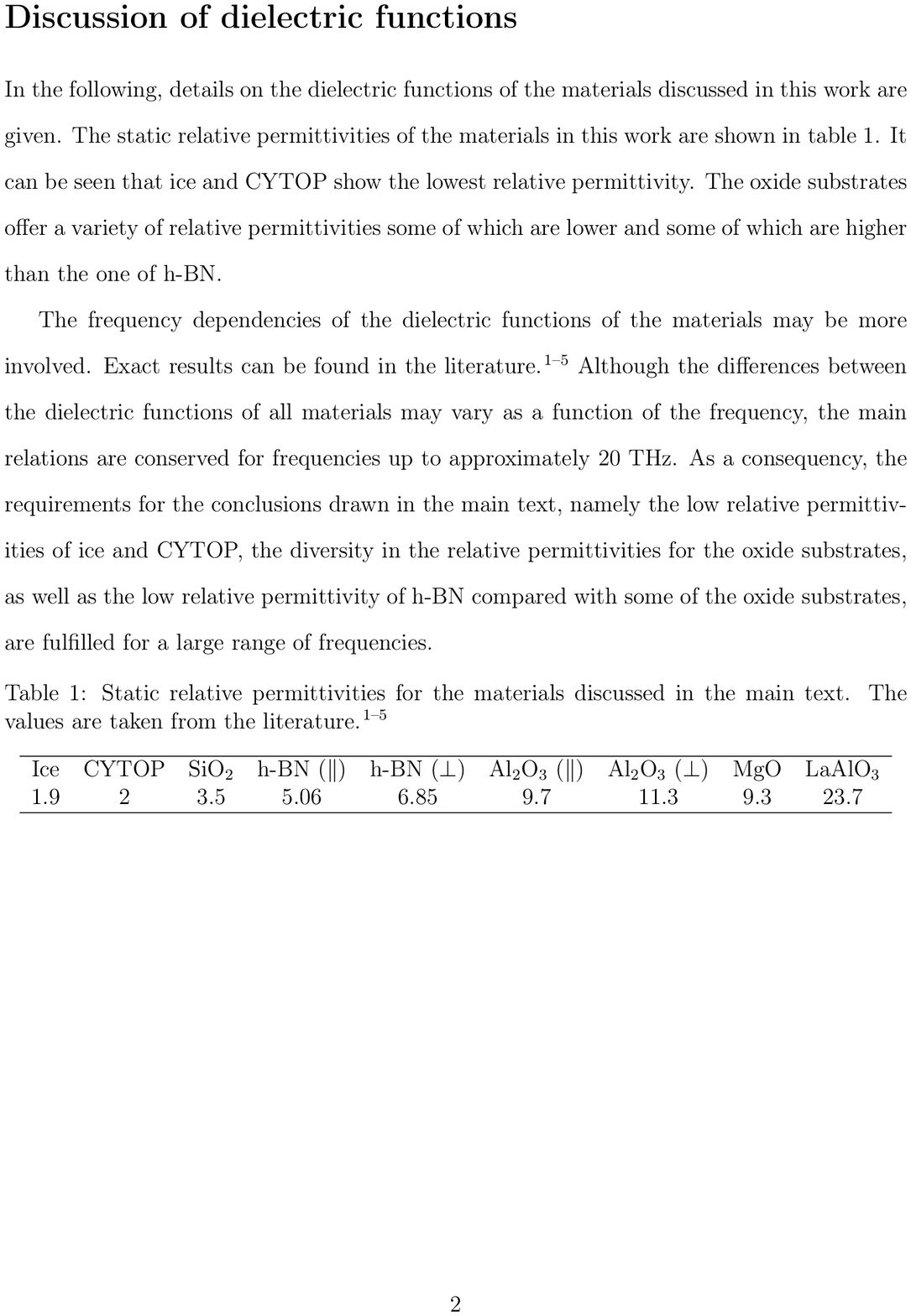}
 \caption{}
\end{figure*}

\newpage 

\begin{figure*}
\centering
 \includegraphics{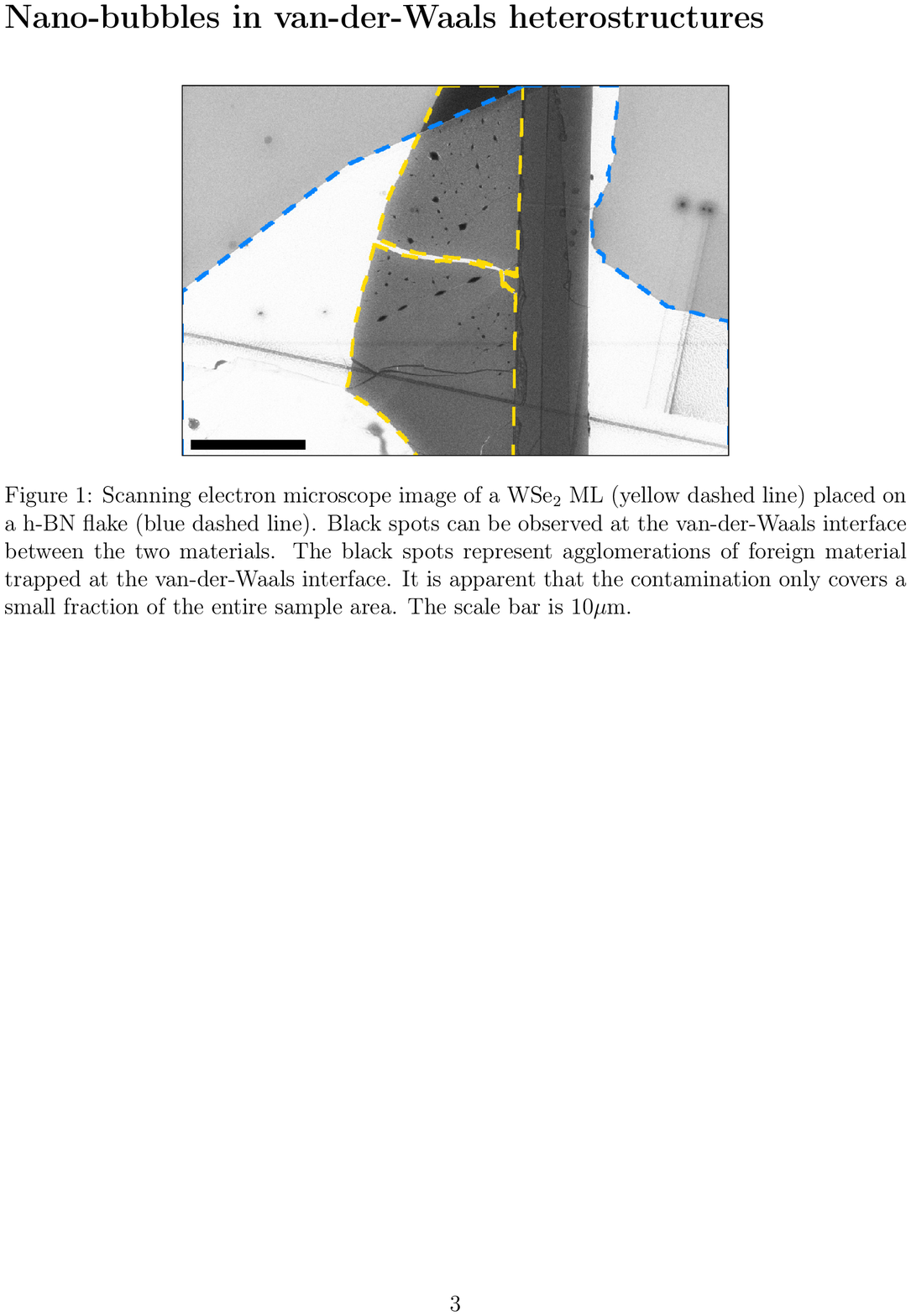}
 \caption{}
\end{figure*}

\newpage 

\begin{figure*}
\centering
 \includegraphics{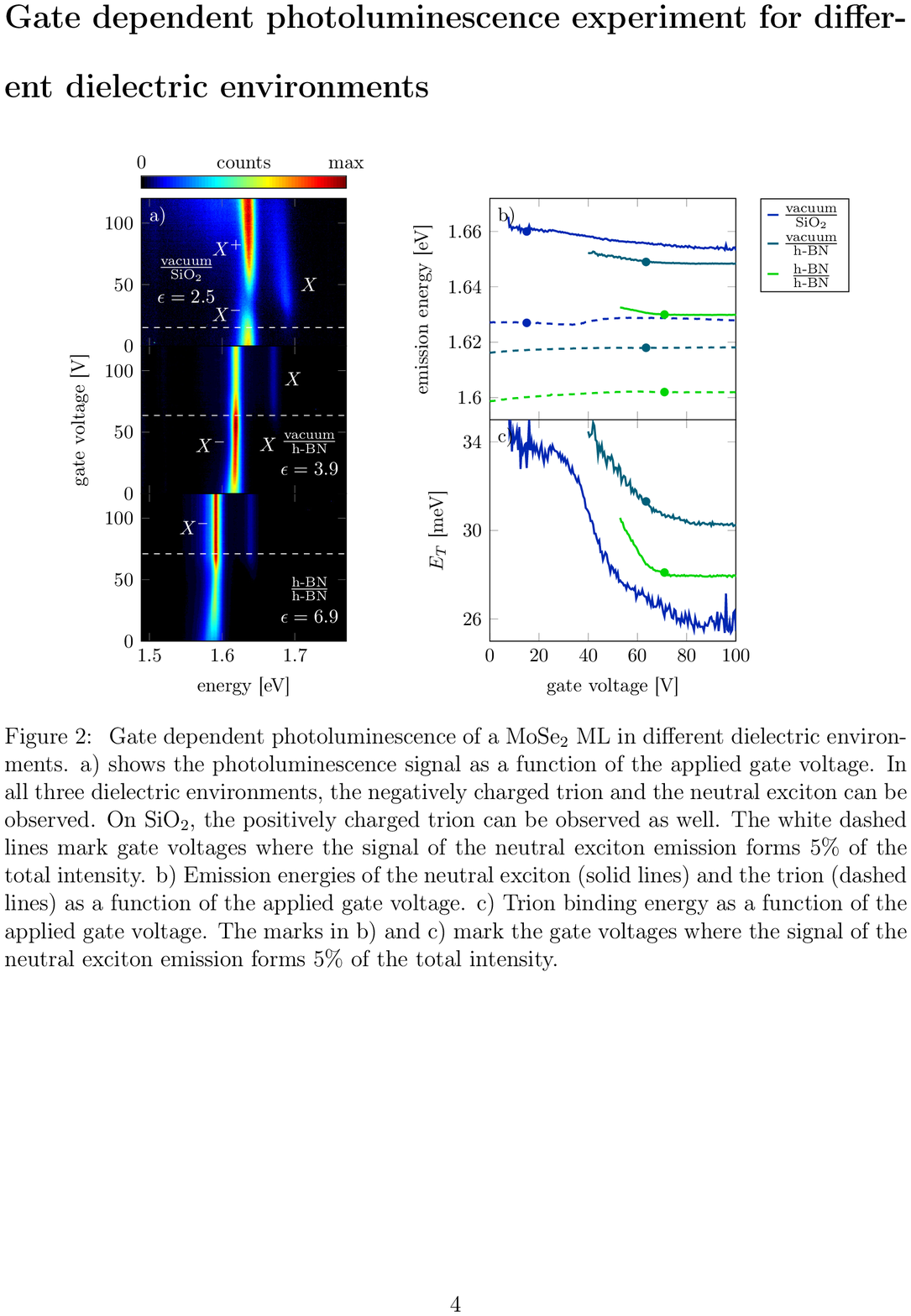}
 \caption{}
\end{figure*}

\newpage 

\begin{figure*}
\centering
 \includegraphics{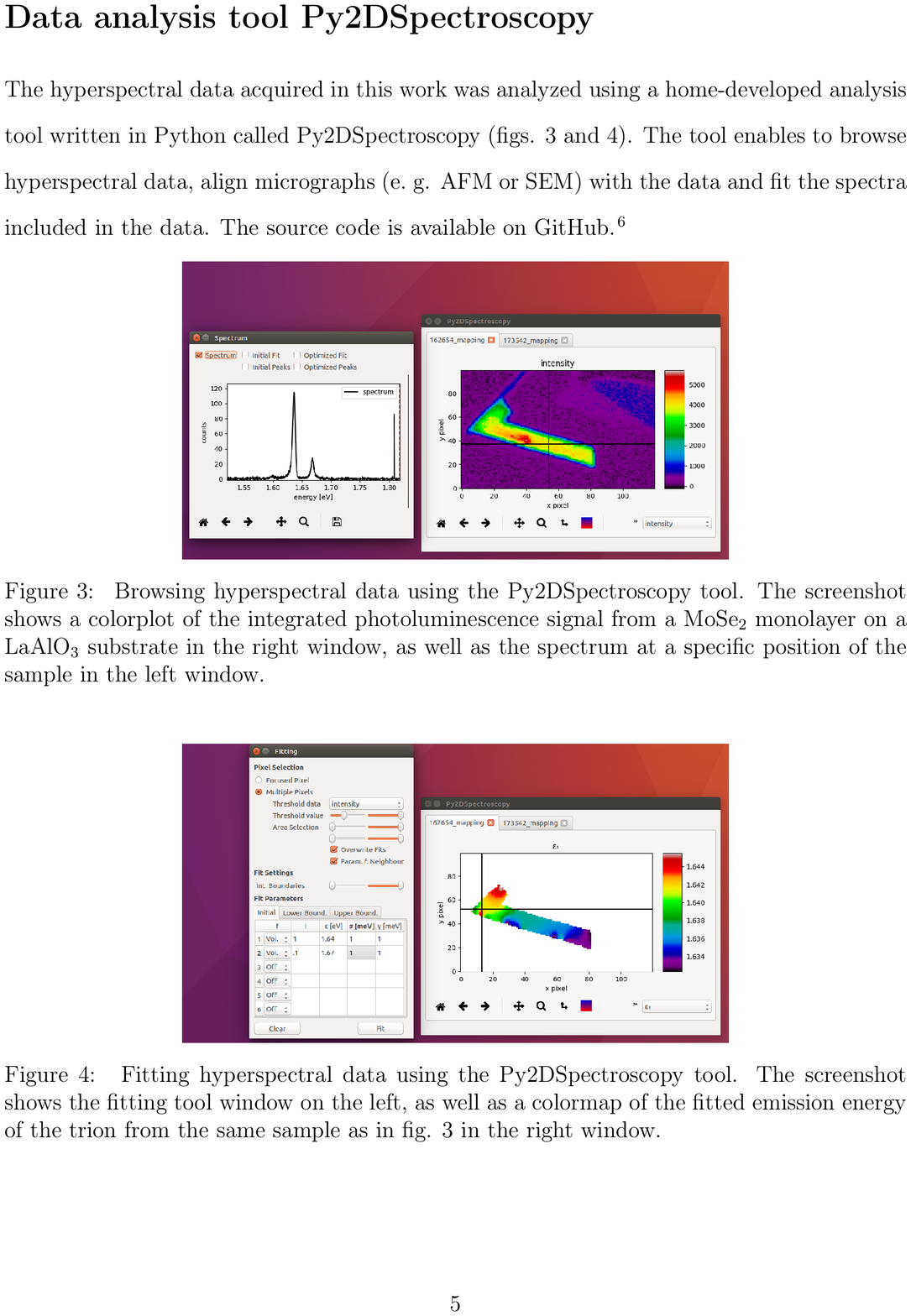}
 \caption{}
\end{figure*}

\newpage 

\begin{figure*}
\centering
 \includegraphics{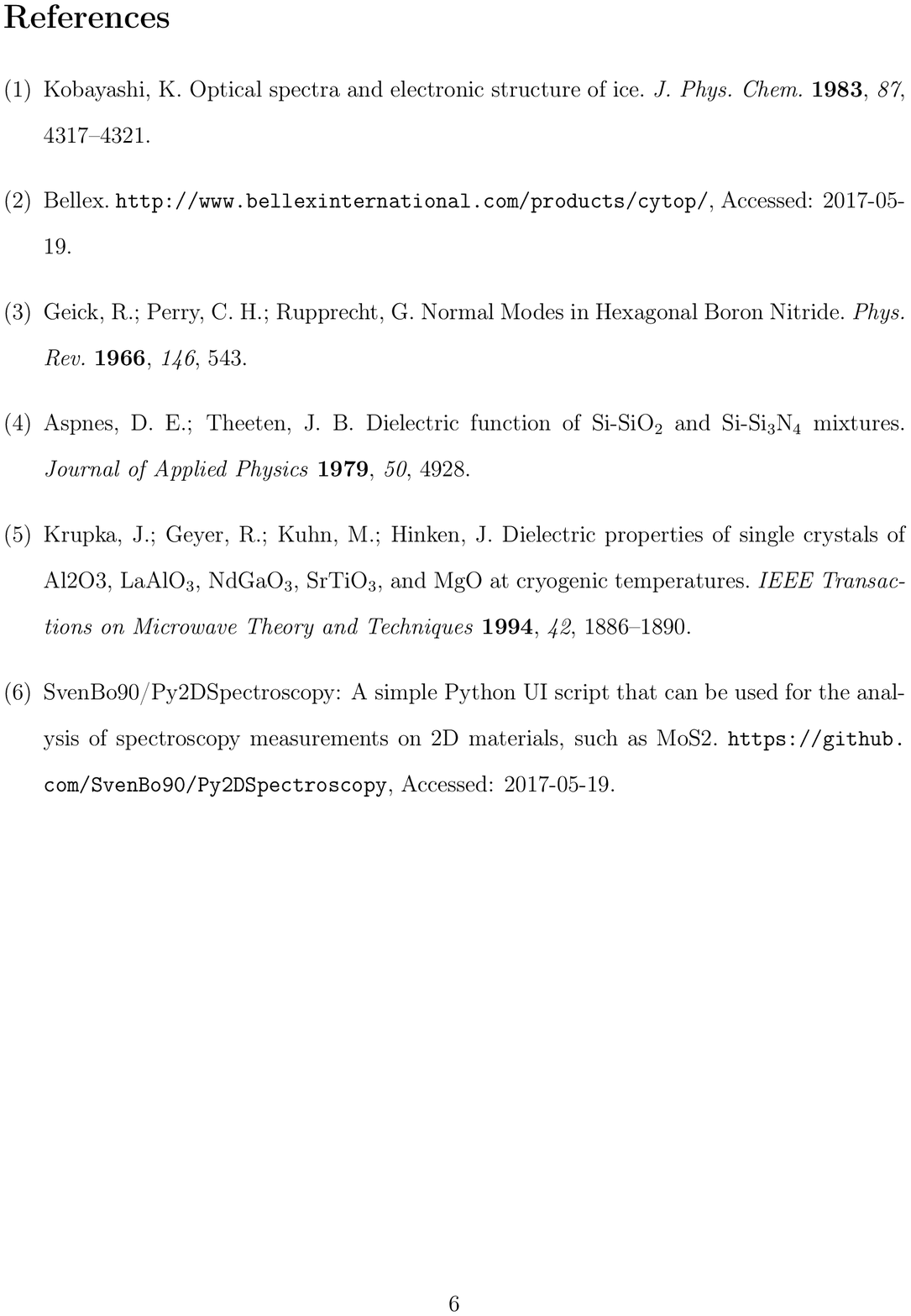}
 \caption{}
\end{figure*}

\end{document}